\documentstyle[12pt]{article}
\catcode`\@=11

\def\@normalsize{\@setsize\normalsize{15pt}\xiipt\@xiipt
\abovedisplayskip 14pt plus3pt minus3pt%
\belowdisplayskip \abovedisplayskip
\abovedisplayshortskip  \z@ plus3pt%
\belowdisplayshortskip  7pt plus3.5pt minus0pt}
\def\small{\@setsize\small{13.6pt}\xipt\@xipt
\abovedisplayskip 13pt plus3pt minus3pt%
\belowdisplayskip \abovedisplayskip
\abovedisplayshortskip  \z@ plus3pt%
\belowdisplayshortskip  7pt plus3.5pt minus0pt
\def\@listi{\parsep 4.5pt plus 2pt minus 1pt
            \itemsep \parsep
            \topsep 9pt plus 3pt minus 3pt}}

\def\underline#1{\relax\ifmmode\@@underline#1\else
        $\@@underline{\hbox{#1}}$\relax\fi}
\@twosidetrue
\relax

\catcode`@=12

\catcode`\@=11

\def\section{\@startsection{section}{1}{\z@}{3.5ex plus 1ex minus
   .2ex}{2.3ex plus .2ex}{\large\bf}}

\def\ps@headings{\def\@oddfoot{}\def\@evenfoot{}
\def\@oddhead{\hbox{}\hfill
        \makebox[.5\textwidth]{\raggedright\ignorespaces --\thepage{}--
        \hfill }}
\def\@evenhead{\@oddhead}
\def\subsectionmark##1{\markboth{##1}{}}
}

\ps@headings

\catcode`\@=12

\relax

\def\figcap{\section*{Figure Captions\markboth
        {FIGURECAPTIONS}{FIGURECAPTIONS}}\list
        {Fig. \arabic{enumi}:\hfill}{\settowidth\labelwidth{Fig. 999:}
        \leftmargin\labelwidth
        \advance\leftmargin\labelsep\usecounter{enumi}}}
 \relax
\def\tablecap{\section*{Table Captions\markboth
        {TABLECAPTIONS}{TABLECAPTIONS}}\list
        {Table \arabic{enumi}:\hfill}{\settowidth\labelwidth{Table 999:}
        \leftmargin\labelwidth
        \advance\leftmargin\labelsep\usecounter{enumi}}}
 \relax
 \relax

\catcode`\@=11

\def\marginnote#1{}
\newcount\hour
\newcount\minute
\newtoks\amorpm
\hour=\time\divide\hour by60
\minute=\time{\multiply\hour by60 \global\advance\minute by-
\hour}
\edef\standardtime{{\ifnum\hour<12 \global\amorpm={am}%
    \else\global\amorpm={pm}\advance\hour by-12 \fi
    \ifnum\hour=0 \hour=12 \fi
    \number\hour:\ifnum\minute<100\fi\number\minute\the\amorpm}}
\edef\militarytime{\number\hour:\ifnum\minute<100\fi\number\minute}
\def\draftlabel#1{{\@bsphack\if@filesw {\let\thepage\relax
  \xdef\@gtempa{\write\@auxout{\string
    \newlabel{#1}{{\@currentlabel}{\thepage}}}}}\@gtempa
    \if@nobreak \ifvmode\nobreak\fi\fi\fi\@esphack}
     \gdef\@eqnlabel{#1}}
\def\@eqnlabel{}
\def\@vacuum{}
\def\draftmarginnote#1{\marginpar{\raggedright\scriptsize\tt#1}}
\def\draft{\oddsidemargin -.5truein
        \def\@oddfoot{\sl preliminary draft \hfil
        \rm\thepage\hfil\sl\today\quad\militarytime}
        \let\@evenfoot\@oddfoot \overfullrule 3pt
        \let\label=\draftlabel
        \let\marginnote=\draftmarginnote
   
\def\@eqnnum{(\theequation)\rlap{\kern\marginparsep\tt\@eqnlabel}%
\global\let\@eqnlabel\@vacuum}  }
\def\preprint{\twocolumn\sloppy\flushbottom\parindent 1em
        \leftmargini 2em\leftmarginv .5em\leftmarginvi .5em
        \oddsidemargin -.5in    \evensidemargin -.5in
        \columnsep 15mm \footheight 0pt
        \textwidth 250mmin      \topmargin  -.4in
        \headheight 12pt \topskip .4in
        \textheight 175mm
        \footskip 0pt
        
\def\@oddhead{\thepage\hfil\addtocounter{page}{1}\thepage}
        \let\@evenhead\@oddhead \def\@oddfoot{} \def\@evenfoot{} 
}
\def\titlepage{\@restonecolfalse\if@twocolumn\@restonecoltrue\onecolumn
     \else \newpage \fi \thispagestyle{empty}\c@page\z@
        \def\thefootnote{\fnsymbol{footnote}} }
\def\endtitlepage{\if@restonecol\twocolumn \else  \fi
        \def\thefootnote{\arabic{footnote}}
        \setcounter{footnote}{0}}  
\catcode`@=12
\relax

\def\ps@headings{\def\@oddfoot{}\def\@evenfoot{}
\def\@oddhead{\hbox{}\hfill
        \makebox[.5\textwidth]{\raggedright\ignorespaces --\thepage{}--
        \hfill }}
\def\@evenhead{\@oddhead}
\def\subsectionmark##1{\markboth{##1}{}}
}

\ps@headings

\relax

\def\firstpage#1#2#3#4#5#6{
\begin{document}
\begin{titlepage}
\nopagebreak
\title{\begin{flushright}
         \vspace*{-3cm}
        {\normalsize CERN-TH/98-157}\\[-5mm]
        {\normalsize hep-th/9805131}\\[5mm]
\end{flushright}
\vspace{2cm}
{#3}}
\author{\large #4 \\[0.0cm] #5}
\maketitle
\vskip 2mm
\nopagebreak 
\begin{abstract}
{\noindent #6}
\end{abstract}
\vfill
\begin{flushleft}
\rule{16.1cm}{0.2mm}\\[-1mm]
$^\ast$e-mail: Alexandros.Kehagias@cern.ch \\[2mm]
May 1998
\end{flushleft}
\thispagestyle{empty}
\end{titlepage}}

\def\simlt{\stackrel{<}{{}_\sim}}
\def\simgt{\stackrel{>}{{}_\sim}}
\newcommand{\dal}{\raisebox{0.085cm}
{\fbox{\rule{0cm}{0.07cm}\,}}}

\newcommand{\be}{\begin{eqnarray}}
\newcommand{\ee}{\end{eqnarray}}
\newcommand{\G}{\Gamma}
\newcommand{\g}{\gamma}
\newcommand{\bb}{\bar{b}}
\newcommand{\ba}{\bar{a}}
\newcommand{\bz}{\bar{z}}
\newcommand{\bt}{\bar{\tau}}
\newcommand{\e}{\epsilon}
\newcommand{\ot}{\otimes}
\newcommand{\p}{\partial}
\newcommand{\btau}{\bar{\tau}}
\newcommand{\bp}{\bar{\partial}}
\newcommand{\cR}{{\cal R}}
\newcommand{\tR}{\tilde{R}}
\newcommand{\tcR}{\tilde{\cal R}}
\newcommand{\hR}{\hat{R}}
\newcommand{\hcR}{\hat{\cal R}}
\newcommand{\oE}{\stackrel{\circ}{E}}
\renewcommand{\p}{\partial}
\renewcommand{\bp}{\bar{\partial}}
\newcommand{\bP}{{\bf P}}
\newcommand{\cn}{{\stackrel{\circ}{\nabla}}}

\newcommand{\gsi}{\,\raisebox{-0.13cm}{$\stackrel{\textstyle
>}{\textstyle\sim}$}\,}
\newcommand{\lsi}{\,\raisebox{-0.13cm}{$\stackrel{\textstyle
<}{\textstyle\sim}$}\,}
\date{}
\firstpage{3118}{IC/95/34}
{\large {\Large N}EW {\Large T}YPE  {\Large IIB}   
{\Large V}ACUA AND {\Large T}HEIR \\   
{\Large F}--{\Large T}HEORY  
{\Large I}NTERPRETATION$^\star$ \\
\phantom{X}}
{A. Kehagias$^\ast$} 
{
\normalsize\sl Theory Division, CERN, 1211 Geneva 23, Switzerland
}
{We discuss a 
D3-D7 system  in type IIB string theory. The near-horizon geometry 
is described by $AdS^5\times X^5$ where $X^5$ 
is a $U(1)$ bundle over a K\"ahler-Einstein complex surface ${\cal S}$ with 
positive first Chern class $c_1>0$. The surface ${\cal S}$ can either be
$\bP^1\times\bP^1$, $\bP^2$ or $\bP_{n_1,...,n_k}$, a 
blow up of $\bP^2$ at k points with $3\leq k\leq 8$. The $\bP^2$ corresponds
to the maximally supersymmetric $AdS^5\times S^5$ vacuum while 
the other cases lead to vacua with less supersymmetries. In the F-theory
context they can be viewed as compactifications on elliptically fibered 
almost Fano 3-folds.}

\newpage


It has recently been argued in \cite{malda} and further been elaborated 
in a series of papers \cite{kleb}-\cite{SW},
that the large $N$ limit of certain conformal field theories (CFT)
can be described in terms of Anti de-Sitter (AdS)  
supergravity. The CFT
lives on the AdS boundary and a precise recipe for expressing
correlation functions of the boundary theory in terms of the bulk theory 
has been given \cite{kleb},\cite{wit1}.    
In particular,  the four-dimensional ${\cal N}=4$ supersymmetric 
$SU(N)$ Yang-Mills theory is described by the type IIB string theory  on 
$AdS_5\times S^5$ where the radius of both 
the $AdS_5$ and  $S^5$ are proportional 
to $N$. In fact, 
the gauge group $SU(N)$ can be replaced by
$SO(N)$ or $Sp(N/2)$ \cite{wit2} by some appropriate orientifold operations
\cite{Kze},\cite{AOY},\cite{FS}. 
Then, the large $N$ limit in field theory corresponds to the 
type IIB supergravity vacuum. The symmetry of the latter  
 is $SO(4,2)\times SU(4)$ which is just the even subgroup of the 
$SU(2,2|4)$ superalgebra.   
In fact, all simple supersymmetry algebras have
been classified in \cite{WN}\footnote {The $AdS_3$ supergroup is missing in
the list of \cite{WN} since the Anti de-Sitter group in three dimensions 
$SO(2,2)=SL(2)\times SL(2)$ is not simple}. From the list in \cite{WN} one 
may see that the $AdS_{p+2}$
 group $SO(p+1,2)$ can be extended to a supergroup, besides $p=4$, for 
$p=2$ and $p=5$ as well. These cases correspond
to the  $AdS_{4,7}\times S^{7,4}$ vacua
of eleven-dimensional supergravity which, consequently, describes large
$N$ boundary CFT in three and six dimensions.

In addition to the ${\cal N}=4$ supersymmetry algebra $SU(2,2|4)$,
there also exist the superalgebras $SU(2,2|2)$ and 
$SU(2,2|1)$. Their  
even subgroups are  $SO(4,2)\times U(2)$ and $SO(4,2)\times U(1)$, 
respectively, and they are realized by  conformal field theories with less 
supersymmetries, namely, ${\cal N}=2$ and ${\cal N}=1$ superconformal 
Yang-Mills theories. Their supergravity interpretation is chiral supergravity
on $AdS_5\times S^5/\Gamma$ where $\Gamma$ is a discrete subgroup with $\Gamma
\subset SU(2)$ and $\Gamma\subset SU(3)$ for the ${\cal N}=2$ and ${\cal N}=1$
cases, respectively \cite{KS}. 
If, in particular, $\Gamma$ is not in $SU(3)$, 
then we get a non-supersymmetry theory.  
However, in all these cases, as has been shown both 
in  string-theory \cite{LNV} and in field theory context \cite{BJ},\cite{Kze} 
correlation functions of the ${\cal N}=0,1,2$ theories constructed 
by orbifolding are the same as those of the ${\cal N}=4$ theory in the 
large $N$ limit. 
Here, we will examine if there exist  
 vacuum type IIB configurations, other than the  $S^5$ orbifolds,
which realize the other supersymmetry algebras, namely     
$SU(2,2|2)$ and  $SU(2,2|1)$.  Since they
always contain in the even part, besides the Anti de-Sitter group, a $U(1)$
factor, the five dimensional space which replaces the $S^5$ is a $U(1)$
bundle over a four-dimensional space which by supersymmetry turns out to be 
a complex surface with  positive first Chern class.


The massless  bosonic spectrum of  the type IIB superstring theory 
consists in  the graviton $g_{MN}$, 
the dilaton $\phi$ and the antisymmetric tensor $B^1_{MN}$ 
in the NS-NS sector, while in the R-R sector it contains  
the axion  $\chi$, the two-form $B^2_{MN}$ and the 
self-dual four-form field $A_{MNPQ}$. 
The fermionic superpartners are a complex Weyl gravitino $\psi_M$ 
$(\gamma^{11}\psi_M=\psi_M)$
and a  complex Weyl dilatino $\lambda$ $(\gamma^{11}\lambda=-\lambda)$.
The theory has two 
supersymmetries generated by two supercharges of the same chirality.   
It has in addition a conserved $U(1)$ 
charge, which generates  rotations of the two 
supersymmetries and under of  which some of the  fields  are charged. 
In particular, 
the graviton and the four-form field are neutral,  the antisymmetric  
tensors have charge  $q=1$, the  scalars have $q=2$, whereas 
the gravitino and the dilatino have charges $q=1/2$ and $q=3/2$, respectively.

The two scalars of 
the theory can be combined into a complex 
one, $\tau=\tau_1+i\tau_2$, defined by
$
\tau=\chi+ie^{-\phi}\, ,  
$
which parametrizes an $SL(2,{\bf R})/U(1)$ coset space.
The theory  has an  $SL(2,{\bf R})$ symmetry that acts as 
\be
&&\tau\rightarrow \frac{a\tau+b}{c\tau+d}\, , ~~~~
B^\alpha_{MN}\rightarrow {(\Lambda^T)^{-1}}^\alpha_{\mbox{\phantom{a}}\beta} 
B^\beta_{MN} 
\, , ~~~ \Lambda= \left(\matrix{a & b 
\cr c& d}\right) \!\in\! SL(2,{\bf R})\, , \nonumber \\
&&g_{MN}\rightarrow g_{MN}\, ,~~~A_{MNPQ}\rightarrow A_{MNPQ}\, , 
\ee
while  the fermions transform accordingly as 
\be
\psi_M\rightarrow \left(\frac{c\bt+d}{c\tau+d}\right)^{1/4}\psi_M,~~~
\lambda\rightarrow
\left(\frac{c\bt+d}{c\tau+d}\right)^{3/4}\lambda. \nonumber
\ee
 
It is quite convenient to  introduce
$SL(2,{\bf R})$-covariant objects.  For this, let us 
recall that the group 
$SL(2,\bf{R})$ can be represented by a matrix $V^\alpha_{\pm}$:
\be
 V=\left(\matrix{V^1_- & V^1_+\cr 
V^2_- & V^2_+ }\right)= \frac{1}{\sqrt{-2 i\tau_2}} 
\left(\matrix{\bt e^{-i\theta} & \tau e^{i\theta} \cr 
e^{-i\theta} & e^{i\theta} }\right)\, . \nonumber 
\ee
 The local $U(1)$ is realized by the shift 
$\theta\rightarrow \theta+\Delta\theta$ and the global 
$SL(2,\bf{R})$ acts from the left. One may define the quantities 
\be
P_M=-\epsilon_{\alpha\beta}V^\alpha_+\p_MV^\beta_+=ie^{2i\theta}
\frac{\p_M\tau}{2\tau_2}\, , ~~~ Q_M=-i\epsilon_{\alpha\beta}V^\alpha_+
\p_MV^\beta_-=\p_M\theta -
\frac{\p_M\tau_1}{2\tau_2}\, ,   
\ee
where $Q_M$ is a composite $U(1)$ gauge connection and $P_M$ has charge 
$q=2$. We also define the complex three-form
\be
G_{KMN}=-\sqrt{2i}\delta_{\alpha\beta}
V^\alpha_+H^\beta_{KMN}=-i\frac{e^{i\theta}}{\sqrt{\tau_2}}
(\tau H^1_{KMN}+H^2_{KMN})\, , \nonumber
\ee
with charge $q=1$ as well as 
the five-form field strength $F_{MNPQR}$ 
\be
F_{MNPQR}=5\partial_{[M}A_{NPQR]}-\frac{5}{4}Im \left((B^1+iB^2)_{MN}
(H^1-iH^2)_{PQR}\right)\, , \nonumber
\ee 
with charge $q=0$.
We  fix the $U(1)$ gauge by choosing  
$\theta\equiv 0$ from now on. 
In this case, the global $SL(2,\bf{R})$ transformation is 
non-linearly realized and the various fields   transform as
\be
&&P_M\rightarrow \frac{c\bt+d}{c\tau+d}P_M\, , ~~ Q_M\rightarrow 
Q_M+\frac{1}{2i}\p_M\ln \left(\frac{c\bt+d}{c\tau+d}\right)\,  ,\nonumber \\
&&G_{KMN}\rightarrow \left(\frac{c\bt+d}{c\tau+d}\right)^{1/2}\!\!G_{KMN}
\, , ~~~ F_{MNPQR}\rightarrow F_{MNPQR}\, . \nonumber
\ee
We also define  the covariant derivative $D_M=\nabla_M-iqQ_M$, 
which transforms under $SL(2,\bf{R})$ as  
$D_M\rightarrow \left(\frac{c\bt+d}{c\tau+d}\right)^{q/2}D_M \, . \nonumber 
$

The supersymmetry transformations of the dilatino and gravitino 
in a pure bosonic background are \cite{Sw} 
\begin{eqnarray}
\delta\lambda&=&i\gamma^MP_M\epsilon^\ast -\frac{i}{24}\gamma^{MNK}G_{MNK}
\epsilon \, , \label{dil1}\\
\delta\psi_M&=&D_M\epsilon
\!+\!\frac{i}{480}
\gamma^{M_1\cdots M_5} F_{M_1\cdots M_5}
\g_M\epsilon
-\frac{1}{96}\left({\gamma_M}^{NKL}
G_{NKL}-9\gamma^{NL}G_{MNL}\right)\epsilon^\ast
\, , \label{grav1}
\end{eqnarray}
where
$$D_M\epsilon= (\partial_M+\frac{1}{4}{\omega_M}^{AB}\gamma_{AB}
-\frac{1}{2}iQ_M)\epsilon,$$
and  $\epsilon$ is a complex Weyl spinor $(\g^{11}
\epsilon=\e)$.
The five-form $F_{MNPQR}$ satisfies 
\be
F_{MNPQR}=-\frac{1}{5!}\varepsilon_{MNPQRSTUVW}F^{STUVW}\, , \label{sd}
\ee
i.e., it is anti-self dual. This can be seen from the supersymmetry 
transformation of the gravitino eq.(\ref{grav1}) since the factor 
$\gamma^{M_1\cdots M_5}\gamma_M\e$ and the chirality of $\e$ projects out the 
self-dual part of $F_{MNPQR}$.

Omitting  fermions, the field equations for the bosonic 
fields of type IIB supergarvity turn out to be
\begin{eqnarray}
D^MP_M&=&\frac{1}{24}G_{MNK}G^{MNK}\, , ~~~~
D^MG_{MNK}=P^MG_{MNK}^\ast-\frac{2}{3}iF_{MNKPQ}G^{MPQ} \, , \nonumber \\
R_{MN}&=& P_MP_N^\ast+P_M^\ast P_N+\frac{1}{6}F_{MKLPQ}{F_N}^{KLPQ} 
\nonumber \\
&&+\frac{1}{8}\left({G_M}^{PQ}G_{NPQ}^\ast +{G_N}^{PQ}G_{MPQ}^\ast-
\frac{1}{6}g_{MN}G_{KPQ}^\ast G^{KPQ}\right)\, . \label{Einst}
\end{eqnarray}
There also exists a number of indentities between the fields appearing  
in eqs.(\ref{dil1},\ref{grav1}) 
\begin{eqnarray}
&&D_{[M}P_{N]}=0\, , ~~~~\partial_{[M}Q_{N]}=-iP_{[M}P_{N]}\, , 
~~~~\partial_{[M}P_{N]}=-2iP_{[M}Q_{N]}\, , \nonumber \\
&&D_{[M}G_{NKL]}=-P_{[M}G^\ast_{NKL]}\, , ~~~~
\partial_{[M}F_{NKLPQ]}=\frac{5}{12}iG_{[MNK}G^\ast_{LPQ]}\, , \nonumber
\end{eqnarray}
and which follow from the definitions of $P_M,Q_M,G_{MNP}$ and $F_{MNPQR}$.


We will now consider supersymmetric backgrounds which preserve 
some supersymmetry in the presence of D7- and D3-branes. Here, the 
non-vanishing bosonic fields are the graviton $g_{MN}$, 
the complex scalar $\tau$ and the  four-form potential
$A_{MNKL}$. Then, the conditions for unbroken 
supersymmetry turn out to be
\begin{eqnarray}
\gamma^MP_M\epsilon&=&0\, , \label{3p} \\
D_M\epsilon+\frac{i}{480}\gamma^{M_1\cdots M_5}F_{M_1\cdots M_5}
\g_M\epsilon&=&0 \, . \label{3grav}
\end{eqnarray} 

We assume that the  
ten-dimensional space-time is of the form $M^4\times B^6$ so that part of the 
D7 is inside $B^6$. We
split the coordinates $x^M$ as $x^M=(x^\mu,x^m)$ where $(\mu=0,\cdots,3\,~
m=1,\cdots,6)$. The $\gamma$-matrices  split  acoordingly as          
\begin{eqnarray}
\gamma^\mu=\Gamma^\mu\otimes 1\, , ~~~~~ 
\gamma^m=\Gamma^5\otimes \G^m\, ,  \label{GG}
\end{eqnarray}
where $\G^\mu,\G^m$ 
are $SO(1,3)$ and $SO(6)$ $\Gamma$-matrices respectively. We also define 
four- and six-dimensional chirality matrices $\Gamma^5, \Gamma^7$ as  
\be
\G^5=i\G^0\cdots\G^3\, , ~~~~~ \G^7=-i\G^4\cdots\G^9 \, ,  \nonumber
\ee
so that $\gamma^{11}=\G^5\G^7$ and $(\G^5)^2=(\G^7)^2=1$. 
In the representation (\ref{GG}), 
$\G^\mu$ are real and hermitian apart from
$\G^0$ which is 
anti-hermitian while $\G^a$ as well as $\G^5$ and $\G^7$ 
are imaginary and hermitian. 
The topology of  space-time  allows a non-zero five-form  field  of the form
\be
F_{\mu\nu\rho\kappa m}=\epsilon_{\mu\nu\rho\kappa} F_m\, , ~~~
F_{mn\ell pq}=\epsilon_{mn\ell pqr}\tilde{F}^r\, , 
\ee
where $F_m,\tilde{F}_m$ are vectors in $B^6$ which depend on $x^m$ only. 
They are not independed since the (anti) self-duality condition 
eq.(\ref{sd}) gives
\be
F_m=\tilde{F}_m\, . \nonumber
\ee 
Moreover, we assume that the complex scalar $\tau$ depence only on the 
$B^6$ coordinates so that $P_M=(0,P_m)$.   
One may easily verify that 
\be
i\gamma^{M_1\cdots M_5}F_{M_1\cdots M_5}=5!\left(\G^5\otimes F^m\G_m\G^7-
1\otimes F^m\G_m
\right)
\, , 
\ee
so that  eqs.(\ref{3p},\ref{3grav}) turns out to be 
\begin{eqnarray}
\Gamma^5\otimes \Gamma^m P_m\e&=&0\, ,\label{30}\\
D_\mu\e+\frac{1}{2}(\G^\mu\otimes F^m\G_m)\e&=&0\, , \label{31} \\
D_m\e-\frac{s}{2}F_m\e+\frac{s}{2}(1\otimes F^n\G_{mn})\e&=&0 
\, , \label{32}
\end{eqnarray}
where $s$ is the four-dimensional chirality of 
$\e$ $\left(\Gamma^5\otimes 1\e=s\e\right)$.
To solve the above equations, 
we  assume that the metric is of the form
\be
ds^2=e^{2A(x^m)}\eta_{\mu\nu}dx^\mu dx^n+e^{2B(x^m)}h_{mn}dx^mdx^n\,  , 
\label{g}
\ee
where $g_{mn}$ is the metric on $B^6$. Then, eqs.(\ref{31},\ref{32}) are 
written as
\be 
\p_\mu\e+\frac{1}{2}\left(s\p_nA-F_n\right) \Gamma_\mu\otimes \G^n\e&=&0\, , 
\label{f1} \\
\nabla_m\e-\frac{s}{2}F_m\e+\frac{1}{2}\left(\p_nB+sF_n\right)
1\otimes{\G_m}^n
\e&=&0\, , \label{f2}
\ee
where $\nabla_m=\p_m+\frac{1}{4}
\omega_{mab}(h)\Gamma^{ab}-\frac{i}{2}Q_m$ 
is the gauge spin-covariant derivative with respect to the  metric $h_{mn}$. 
By splitting the spinor $\e$ as $\e=e^{A/2}\theta\otimes\eta$, 
 with $\Gamma^5\theta=\theta,~~\Gamma^7\eta=\eta$,
eqs.(\ref{f1}, \ref{f2}) give
\be
F_m=\p_m A\, , ~~~~~B=-A\, ,
\ee
and eq.(\ref{f2}) is then reduced to 
\be
\nabla_m\eta=0\, . \label{eett}
\ee
Thus, the number of unbroken supersymmetries is determined by the number 
of gauge-covariantly constant spinors $\eta$.
The integrability condition of eq.(\ref{eett}) is 
\be
R_{mn}(h)=P_m^\ast P_n+P_mP_n^\ast \, , 
\label{equat}
\ee
which combined with the field equations 
\be
R_{MN}=P_MP_N^\ast+P_NP_M^\ast + \frac{1}{96}{F_M}^{KLPQ}F_{NKLPQ} \, ,
\nonumber
\ee
gives that  $A(x^m)$ is harmonic \cite{DP}, 
\be
\frac{1}{\sqrt{h}}\p_m\left(\sqrt{h}h^{mn}\p_n e^{-4A}\right)=0\, .\label{harm}
\ee 
Thus, finally, what remain to be solved 
are eqs.(\ref{eett},\ref{harm}) and the supersymmetric condition
\be
\Gamma^mP_m\eta=0\, . \label{PP}
\ee
We will assume now that the metric $h_{mn}$ takes the form
\be
h_{mn}dx^mdx^n=dr^2+r^2g_{ij}dx^idx^j \, , ~~~~(i,j=1,...,5) \, , \label{M}
\ee
where $g_{ij}$ is the metric of a five-dimensional compact space $X^5$.
In this case, eq.(\ref{harm}) gives
\be
e^{-4A}=1+\frac{Q}{r^4}\, , \nonumber
\ee
where $Q=4\pi g_s N\alpha'$ and the ten-dimensional metric takes the form
\be
ds^2=\left(1+\frac{Q}{r^4}\right)^{-1/2}\left(-dt^2+dx_1^2+dx_2^2+dx_3^2
\right)+\left(1+\frac{Q}{r^4}\right)^{1/2}\left(dr^2+r^2g_{ij}dx^idx^j
\right)\, . 
\ee
This is reduced to  the standard D3-brane solution \cite{HSt} if $X^5$ is 
$S^5$, i.e., when the metric (\ref{M}) is flat. 
The near-horizon geometry at $r\to 0$ turns out to be 
\be
ds_h^2= \frac{r^2}{\sqrt{Q}}\left(-dt^2+dx_1^2+dx_2^2+dx_3^2
\right)+\frac{\sqrt{Q}}{r^2}dr^2 +\sqrt{Q}g_{ij}dx^idx^j\, , 
\ee
and, thus, it is $AdS^5\times X^5$.

Let us now return to the gauge-covariant Killing spinor equation (\ref{eett})
which will determine the possible  spaces $X^5$. 
With the metric (\ref{M}), eq.(\ref{eett}) turns out to be
\be
\left(\partial_r-\frac{i}{2}Q_r\right)\eta=0\, , 
~~~\cn_i\eta+\frac{1}{4}{\Gamma_i}^r\eta
=0\, , \label{spin} 
\ee
where $\cn_i=\p_i+\frac{1}{4}
\omega_{ijk}(h)\Gamma^{jk}-\frac{i}{2}Q_i$ is the  gauge-covariant 
derivative on $X^5$.  
Since the metric (\ref{M}) satisfies   
\be
R_{rr}(h)=R_{ri}=0\, , ~~~R_{ij}(h)=R_{ij}(g)-4g_{ij}\, , 
\ee
we get 
the integrability condition of eq.(\ref{spin})   
\be
0&=& P_r^*P_r=P_r^*P_i+P_rP_i^*\, , \nonumber \\
R_{ij}(g)&=&P_i^\ast P_j+P_iP_j^\ast+4g_{ij}\, .  \label{int}
\ee
Thus, $P_r=0$ and  the complex 
scalar $\tau$ is independent of $r$.
We may split now the $\gamma$-matrices $\Gamma^i$ as 
\be
\Gamma^i=\Sigma^i\otimes\sigma^1\, , 
~~~~\Gamma^r=1\otimes\sigma^3\, , \nonumber 
\ee
where $\Sigma^i$ are $SO(5)$ $\gamma$ matrices. 
The spinor $\eta$ split accordingly as $\eta=\eta_0\otimes\xi$. 
Then, eq.(\ref{spin}) turns out to be 
\be
\cn_i\eta_0=\mp\frac{1}{2}\Sigma_i\eta_0\, , \label{KS}
\ee
so that $\eta_0$ is a Killing spinor on $X^5$. The number of independent 
solutions to eq.(\ref{KS}) specifies the number of unbroken symmetries. 

The Killing spinor equation (\ref{KS}) has extensively been studied 
in the Kaluza-Klein supergravity context. There, the isometries of $X^5$ 
appear as the gauge group in four dimensions. In our case, the isometries of
$X^5$ will lead to gauged supergravities in $AdS^5$ and in view of 
the AdS/CFT correspondance will appear as global symmetries in the boundary 
CFT. Since, the supersymmetric boundary CFT will have at least a global $U(1)$
corresponding to the R-symmetry of the minimal ${\cal N}=1$ case, $X^5$ will
necesserily
have a $U(1)$ isometry. It is then natural to assume that $X^5$ is 
a $U(1)$ bundle over a four-dimensional space with metric of the form 
\be
 g_{ij}dx^idx^j=g_{ab}dx^adx^b+4(d\psi+A_adx^a)^2\, , ~~~(a,b=1,...,4) \, .
\label{mm}
\ee
If the four-dimensional space is a complex K\"ahler 
 surface ${\cal S}$ with metric $g_{ab}$, then $A_adx^a$ is the $U(1)$ 
connection with field strength proportional 
to the K\"ahler two-form of the base ${\cal S}$, $J_{ab}$, i.e., 
\be
F_{ab}=iJ_{ab}\, .
\ee
In this case, the Killing 
spinor equation (\ref{KS}) can be   solved as  
in the seven-dimensional case considered in \cite{PW}.
In addition, the condition eq.(\ref{PP})
gives that 
\be
\Gamma^aP_a\eta+\Gamma^\psi P_\psi\eta=0\, , \label{pssi}
\ee
where $x^i=(x^a,\psi), a=1,...,4$.   By using an explicit representation for 
the $\gamma$-matrices $(\Gamma^a,\Gamma^\psi)$, one may verify that 
eq.(\ref{pssi}) is satisfied if $\tau$ is a holomorphic function   
of the complex coordinates $z^1=x^1+ix^2,~~z^2=x^3+ix^4$ and independent of 
$\psi$.
With the metric (\ref{mm}), the integrability conditions eq.(\ref{int})
are then reduced to 
\be
R_{ab}&=&P_a^\ast P_b+P_aP_b^\ast+6g_{ab}\, . \label{RR}
\ee
In the absence of D7-branes, $(P_a=0)$ we have that 
\be
R_{ab}=6g_{ab}\, , \label{E}
\ee
 and the 
obvious solution is then $\bP^2$, the complex projective space. 
In that case,
$X^5$ is a $U(1)$ bundle over $\bP^2$, which is just the five-sphere $S^5$
and we recover the standard D3-brane solution. However,
there are other solutions as well. Namely, every complex compact surface 
${\cal S}$ which 
satisfies eq.(\ref{E}) provides a solutions as well. Such surfaces have 
positive first Cern-class $c_1>0$ as opposed to the CY's
which have vanishing $c_1$. Surfaces with  $c_1>0$ are known as del 
Pezzo surfaces. These include $\bP^1\times\bP^1$,
or $\bP_{n_1...n_k}^2$, the surface which is obtained by blowing up $\bP^2$
at k generic points (no three-points are colinear and no six points are in 
one quadratic curve in $\bP^2$) with $0\leq k\leq 8$. Thus, the solutions 
to eq.(\ref{E}) is reduced to find all del Pezzo surfaces which admit 
K\"ahler-Einstein metrics \cite{ALB}-\cite{T1}. A complex surface now  
admits a K\"ahler-Einstein metric if and only if its group of automorphism 
is reductive \cite{ALB},\cite{AF}. Surfaces which admit K\"ahler-Einstein 
metrics are for example Fermat cubics in $\bP^3$ \cite{YTS}.   
As have been proven in \cite{TY}, del Pezzo surfaces which 
admit K\"ahler-Einstein metrics are  $\bP^2,~\bP^1\times\bP^1$ and 
$\bP_{n_1...n_k}^2$ with $3\leq k\leq 8$. They 
correspond to boundary CFT with 
${\cal N}=4,2,1$ supersymmetries.

We may also study the above compactifications in the F-theory context 
\cite{Vafa}. Here,  
the complex scalar $\tau$ is identified with the 
complex sctructure moduli of an internal torus. The $SL(2,{\bf Z})$ 
symmetry of type IIB string theory is then geometrically realized as the 
modular group of the torus. The modulus of the torus will vary holomorphically
on the surface ${\cal S}$ defining an elliptically fibed space which then 
will be an almost Fano 3-fold.  For the $\bP^1\times\bP^1$ case for example, 
let us consider the cubic in $\bP^2$
\be
v^3+u^3+w^3+zvuw=0\, . \label{cubic}
\ee 
If $z\in \bP^1$, then the above equation defines the del Pezzo surface 
\be
dP_9=\left[ \begin{array}{c} 
\bP^2 \\ \bP^1 
 \end{array} \right|\left.\begin{array}{c} 3\\1\end{array}\right]\, , 
\ee
a hypersurface of bidegree (3,1) in $\bP^2\times\bP^1$ 
which is clearly elliptically fibered over $\bP^1$. Details for this 
surface can be found in \cite{DW}, \cite{CL}. By a change of variables, 
one may bring eq.(\ref{cubic}) into the Weierstrass form 
\be
y^2=x^3+f_4(z)x+f_6(z)\, , \label{qubic}
\ee
where $f_q(z)$ is a polynomial of degree $q$.  
Then the torus will degenerate at the points where the discriminant 
\be
\Delta=4{f_4}^3+27{f_6}^2\, , 
\ee
vanish. Thus, there are twelve points where the torus degenerates 
which is actually the Euler number of $dP_9$. Then, the three-folds
\be 
Y_1=dP_9\times \bP^1 ~~~~Y_2=\left[ \begin{array}{c} 
\bP^2 \\ \bP^2 
 \end{array} \right|\left.\begin{array}{c} 3\\2\end{array}\right]\, ,
\ee
provide  F-theory compactification on the  almost  Fano three-folds $Y_1,Y_2$.
In both cases, the base of the elliptic fibration is $\bP^1\times\bP^1$. 
However, in the case of $Y_1$, the torus varies only on one $\bP^1$ factor
while in the $Y_2$ case, varies on both.


\end{document}